%% file: specialgeom_pos.tex
\newcommand{\version}{January 27, 2011}
\documentclass[cits,a4paper]{PoS}

\usepackage{amsmath,amsfonts,amssymb,slashed}

\IfFileExists{dsfont.sty}
	{\usepackage{dsfont}
         \newcommand{\id}{\mathds{1}}}
	{\typeout{Package dsfont.sty was not found, using alternative macros.}
         \let\mathds=\mathbb
         \newcommand{\id}{\mbox{1 \kern-.59em {\rm l}}}}

\newcommand{\nocontentsline}[3]{}
\newcommand{\tocless}[3]{\bgroup\let\addcontentsline=\nocontentsline#1{#2}#3\egroup}

\newcommand{\qed}{\nobreak \ifvmode \relax \else
      \ifdim\lastskip<1.5em \hskip-\lastskip
      \hskip1.5em plus0em minus0.5em \fi \nobreak
      \vrule height0.75em width0.5em depth0.25em\fi}

\newcommand{\be}{\begin{equation}}
\newcommand{\ee}{\end{equation}}

\def\nn{\nonumber}
\def\bea{\begin{eqnarray}}
\def\eea{\end{eqnarray}}

\def\beqa{\begin{eqnarray}} 
\def\eeqa{\end{eqnarray}} 
\def\beq{\begin{equation}} 
\def\eeq{\end{equation}}

\def\Tr{{\rm Tr}}

\def\a{\alpha}          
\def\b{\beta}           

\def\r{\rho}
\def\cM{{\cal M}}  \def\cO{{\cal O}}
\def\cP{{\cal P}}

\newcommand{\R}{\mathds{R}}

\def\bit{\begin{itemize}}
\def\eit{\end{itemize}}

\def\({\left(}
\def\){\right)}
\def\diag{\mbox{diag}}

\def\pa{\partial} \def\del{\partial}

\def\bcomment#1{}

\newcommand{\nc}{non-com\-mu\-ta\-tive}

\newcommand{\eqnref}[1]{Eqn.~(\ref{#1})}		% for equations with preceding Eqn.
\newcommand{\figref}[1]{Fig.~\ref{#1}}			% for figures
\newcommand{\inv}[1]{\frac{1}{#1}}				% inverse of something

\newcommand{\syf}{\varTheta}

\newcommand{\nabg}{\nabla^g}
\newcommand{\nabG}{\nabla^G}

\newcommand{\pb}[2]{\{#1,#2\}_{pb}}					% Poisson bracket
\newcommand{\co}[2]{[#1,#2]}						% commutator
\newcommand{\aco}[2]{[#1,#2]_+}						% anticommutator
\newcommand{\starco}[2]{\left[ #1\stackrel{\star}{,}#2\right] }		% star commutator
\newcommand{\intx}{\int\!\! d^4x}					% 4-dim x integral
	
\newcommand{\intg}{\int\!d^4x\sqrt{g}\,}				% integral with measure \sqrt{g}
\newcommand{\ri}{\textrm{i}}
\newcommand{\lpa}{\overleftarrow{\pa}}					% \partial with arrow on to (left or right)
\newcommand{\rpa}{\overrightarrow{\pa}}
\newcommand{\bs}{{\bar{\sigma}}}	% sigma-bar
\newcommand{\Tt}{\tilde{t}}

\renewcommand{\a}{\alpha}
\renewcommand{\b}{\beta}
\newcommand{\g}{\gamma}
\renewcommand{\d}{\delta}
\newcommand{\e}{\epsilon}

\renewcommand{\th}{\theta}

\newcommand{\vth}{\vartheta}

\newcommand{\m}{\mu}
\newcommand{\n}{\nu}

\renewcommand{\r}{\rho}
\newcommand{\s}{\sigma}

\newcommand{\vph}{\varphi}

\newcommand{\w}{\omega}
\renewcommand{\Xi}{\Xi}

\newcommand{\W}{\Omega}
\title{Special Geometries Emerging from Yang-Mills Type Matrix Models}

\ShortTitle{Geometries Emerging from Matrix Models}

\author{\speaker{Daniel N. Blaschke}\\
        University of Vienna, Faculty of Physics\\
        Boltzmanngasse 5, A-1090 Vienna (Austria)\\
        E-mail: \email{daniel.blaschke@univie.ac.at}}

\abstract{I review some recent results which demonstrate how various geometries, such as Schwarzschild and Reissner-Nordstr{\"o}m, can emerge from Yang-Mills type matrix models with branes. 
Furthermore, explicit embeddings of these branes as well as appropriate Poisson structures and star-products which determine the non-commutativity of space-time are provided. 
These structures are motivated by higher order terms in the effective matrix model action which semi-classically lead to an Einstein-Hilbert type action.}

\FullConference{Corfu Summer Institute on Elementary Particles and Physics - Workshop on Non Commutative Field Theory and Gravity,\\
September 8-12, 2010\\
Corfu Greece}

\date{\version}
\graphicspath{{./figures/}}

\begin{document}

\input{specialgeom}

%%%%%%%%%%%%%%%%%%%%%%%%%%%%%%%%%%%%%%%%%%%%%%%%%%%%%%%
\bibliographystyle{../../custom1.bst}
\bibliography{../../articles.bib,../../books.bib}

\end{document}

%% file: specialgeom.tex
\section{Background}
%%%%%%%%%%%%%%%%%%%%%%%%%%%%%%%
In past years, various approaches to quantum space-time have been pursued. One possibility is to replace classical space-time by a {\nc} one where the coordinate functions $x^\m$ are promoted to Hermitian operators $X^\m$ on a Hilbert space $\mathcal{H}$. These ``coordinate'' operators satisfy certain non-trivial commutation relations
\begin{align}
\co{X^\m}{X^\n}=\ri\th^{\m\n} \,,
\end{align}
which in the simplest case reduce to a Heisenberg algebra, i.e. with constant $\th^{\m\n}$. For a review of such {\nc} field theories see e.g.~\cite{Douglas:2001,Szabo:2001,Rivasseau:2007a,Blaschke:2010kw}.
In order to incorporate gravity in this context, however, a dynamical non-constant commutator $\th^{\m\n}$ is required, which semi-classically determines a Poisson structure on space-time. Incidentally, matrix models of Yang-Mills type\footnote{In fact, a supersymmetric version, the 10-dimensional IKKT model~\cite{Ishibashi:1996xs}, is expected to be UV finite and hence might represent a candidate for some form of quantum gravity coupled to matter~\cite{Jack:2001cr,Steinacker:2010rh,Steinacker:2011yx}.} naturally realize this idea --- for a review, see~\cite{Steinacker:2010rh} and~\cite{Steinacker:2007dq,Steinacker:2008,Steinacker:2008a}.
Our starting point is hence the matrix model action
\begin{align}
S_{YM}&=-\Tr\co{X^a}{X^b}\co{X^c}{X^d}\eta_{ac}\eta_{bd}\,,
\label{YM--model}
\end{align}
where $\eta_{ab}$ denotes the flat metric of a $D$-dimensional embedding space with arbitrary signature and $X^a$ are Hermitian matrices on $\mathcal{H}$ which in the semi-classical limit are interpreted as coordinate functions. If one considers some of the coordinates to be functions of the remaining ones~\cite{Steinacker:2008ri}
such that $X^a \sim x^a = (x^\mu,\phi^i(x^\mu))$ in the semi-classical limit,
one can interpret the $x^a$ as defining the embedding of a $2n$-dimensional submanifold 
$\cM^{2n}\hookrightarrow\R^D$ equipped with a non-trivial induced metric
\begin{align}
g_{\m\n}(x)&=\pa_\m x^a \pa_\n x^b\eta_{ab}
=\eta_{\m\n}+\pa_i\phi^i(x)\pa_j\phi^j(x)
\,, 
\end{align}
via pull-back of $\eta_{ab}$, and where $\m,\n\in{1,\ldots,2n}$ and $i,j\in{2n+1,\ldots,D}$. 
Here we consider this submanifold to be a four dimensional space-time $\cM^4$, and following~\cite{Steinacker:2008ri} we can interpret
\begin{align}
-\ri\co{X^\m}{X^\n}\sim \pb{x^\m}{x^\n}= \ri\th^{\m\n}(x)
\end{align}
as a Poisson structure on $\cM^4$. Furthermore, we assume that $\th^{\m\n}$ is non-degenerate, 
so that its inverse matrix $\th^{-1}_{\m\n}$ defines a symplectic form 
\begin{align}
\varTheta=\inv{2}\theta^{-1}_{\mu\nu} dx^\mu \wedge dx^\nu
\end{align}
on $\cM^4$.

However, it is not the induced metric which is ``seen'' by scalar fields, gauge fields, etc., but the effective metric~\cite{Steinacker:2007dq}
\begin{align}
G^{\m\n}&=e^{-\s}\th^{\m\r}\th^{\n\s}g_{\r\s}
\,, &
e^{-\s}&\equiv \frac{\sqrt{\det\th^{-1}_{\m\n}}}{\sqrt{\det G_{\r\s}}}\,.
\end{align}
Therefore, an interesting special case where $G_{\m\n}=g_{\m\n}$ may be considered. In fact, this corresponds to having a (anti-) self-dual symplectic form, i.e. $\star\varTheta=\pm\ri\varTheta$.
This case, however is restricted to 4-dimensional submanifolds $\cM^4$, as in four dimensions one always has $|G|=|g|$ which makes the assumptions above possible. (For details, see~\cite{Steinacker:2010rh}.) 

Let us consider the following example in order to make the effective geometry clearer: The gauge invariant kinetic term of a test particle modelled by a scalar field $\phi$ has the form
\begin{align}
S[\phi]&=-\Tr\co{X^a}{\Phi}\co{X^b}{\Phi}\eta_{ab} \nn\\
&\sim\intx\sqrt{\det\th^{-1}}\th^{\m\n}\pa_\m x^a\pa_\n\phi\th^{\r\s}\pa_\r x^b\pa_\s\phi\eta_{ab} \nn\\
&=\intx\sqrt{\det\th^{-1}}G^{\n\s}\pa_\n\phi\pa_\s\phi
\,.
\end{align}

\section{Curvature}
%%%%%%%%%%%%%%%%%%%%%%%%%%%%%%%
\begin{figure}[ht]
\centering
\includegraphics[scale=1.0]{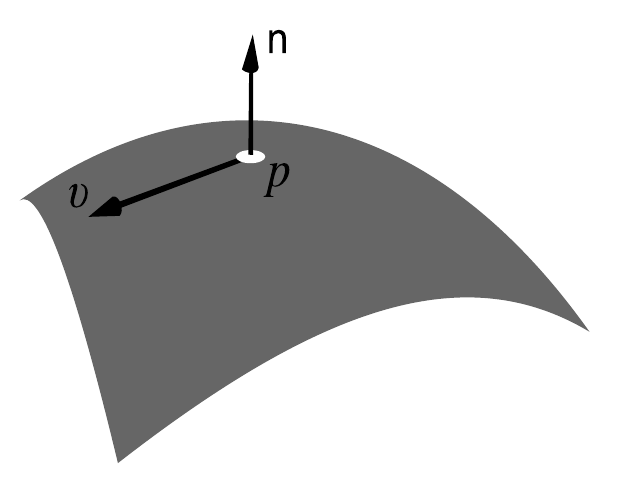}
{\setlength{\unitlength}{1pt}
\put(-145,40){$g_{\mu\nu}(x)$}
\put(-70,40){$\cM^{2n}$}
}
\caption{Embedding and induced metric}
\label{fig:embedding}
\end{figure}
The bare matrix model  \eqnref{YM--model} without matter
leads to the following e.o.m. for $X^c$:
\begin{align}\label{eq:basic-matrix-eom}
\co{X^a}{\co{X^b}{X^c}}\eta_{ab}&=0\,.
\end{align}
Furthermore, one can derive the matrix energy-momentum tensor $T^{ab}$ which reads~\cite{Blaschke:2010rg,Blaschke:2010qj}
\begin{align}
T^{ab}&= H^{ab} - \frac 14\eta^{ab} H\, 
\sim\,\, e^\s\left(\frac{(Gg)}{4} \eta^{ab}  - G^{\mu\nu}\del_\mu x^a \del_\nu x^b \right) \,, \nn\\
H^{ab} &= \inv{2}\aco{\co{X^a}{X^c}}{\co{X^b}{X_c}} %\,,  \qquad 
\sim\,\, -e^\sigma G^{\mu\nu}\del_\mu x^a \del_\nu x^b 
\,, \nn\\
H &=  H^{ab} \eta_{ab}
\,,
\end{align}
and whose conservation follows directly from the matrix equations of motion \eqref{eq:basic-matrix-eom} above:
\begin{align}\label{eq:basic-ward-id}
\co{X^a}{T^{a'b}}\eta_{aa'}=0\,.
\end{align}
Interestingly, there is a close connection between the matrix energy-momentum tensor and the projectors on the tangential/normal bundle of $\cM\in\R^D$
\begin{align}
\cP^{ab}_T&=g^{\m\n}\pa_\m x^a\pa_\n x^b \,,
&\cP^{ab}_N&=\eta^{ab}-\cP^{ab}_T
\,.
\end{align}
Namely, in the special case where both metrics coincide, i.e. the self-dual case where $G_{\m\n}=g_{\m\n}$, 
one has $T^{ab} \sim e^\sigma \cP_N^{ab}$ and $H^{ab} \sim -e^\sigma \cP_T^{ab}$ in the semi-classical limit. 

Furthermore, one easily derives the relation $\nabg_\s\nabg_\n x^a=\cP^{ab}_N\nabG_\s\nabG_\n x_b$, where $\nabg/\nabG$ are the covariant derivatives defined with standard Christoffel symbols with respect to $g/G$, respectively. 
Hence the curvature tensor with respect to the induced metric $g$ can be written as
\begin{align}
R_{\rho\sigma\nu\mu}[g]
&= \nabg_\sigma\nabg_\mu x^a\nabg_\rho\nabg_\nu x_a 
- \nabg_\sigma\nabg_\nu x^a\nabg_\mu\nabg_\rho x_a  \nn\\
&= \cP_N^{ab}\nabG_\sigma\nabG_\mu x_a\nabG_\rho\nabG_\nu x_b
- \cP_N^{ab}\nabG_\sigma\nabG_\nu x_a\nabG_\mu\nabG_\rho x_b 
\,,
\end{align}
where the first line is simply the Gauss-Codazzi theorem, and Latin indices were pulled down with the embedding metric $\eta_{ab}$. Using the tensor $C_{\a;\m\n}:=\pa_\a x^a\nabG_\m\pa_\n x_a$ allows to relate the curvature tensors associated with $G/g$:
\begin{align}
R_{\rho\sigma\nu\mu}[g] &= (Gg)^{\eta}_{\mu} R_{\rho\sigma\nu\eta}[G] +\nabG_\s C_{\m;\r\n} -\nabG_\r C_{\m;\s\n} -C_{\a;\s\m} C_{\b;\r\n} g^{\a\b} +C_{\a;\s\n} C_{\b;\m\r} g^{\a\b} 
\,. 
\end{align}

It was previously shown in~\cite{Blaschke:2010rg,Blaschke:2010qj}, that the Einstein-Hilbert action emerges in the effective matrix model action. In particular, a certain combination of order 10 matrix terms semi-classically leads to
\begin{align*}
S_{\cO(X^{10})}\sim&\,\intx\frac{\sqrt{g}}{(2\pi)^2}e^{2\s}(R[g] - 3 R^{\mu\nu}[g] h_{\mu\nu} ) + \cO(\pa h^2) 
\,, 
\end{align*}
where $G_{\m\n}=g_{\m\n}+h_{\m\n}$ is almost self-dual. 
In the self-dual case (i.e. $h=0$), this reduces to
\begin{align}
S_{\cO(X^{10})}= \Tr\left(2T^{ab}\Box X_a \Box X_b - T^{ab}\Box H_{ab}\right) 
\sim -2 \!\intg e^{2\s}R 
\,, 
\end{align}
where $\Box Y\equiv \co{X^a}{\co{X_a}{Y}}$, 
and additionally one finds the order 6 matrix terms
\begin{align}
S_{\cO(X^{6})}&= \Tr \left( \frac{1}{2} [X^c,[X^a,X^b]] [X_c,[X_a,X_b]] - \Box X^a \Box X_a\right) \nn\\
&\sim \intg  \bigg(\frac 12\theta^{\mu\rho} \theta^{\eta\a} R_{\mu \rho\eta\a} 
- 2 e^\s R + 2 e^\s \del^{\mu}\s \del_\mu\s \!\bigg) .
\end{align}

In general, however, the degrees of freedom are given by the embedding $\phi^i$ and the deviation from the self-dual Poisson structure $A_\m$, i.e.:
\begin{align}
\th^{-1}_{\m\n} &= \bar\th^{-1}_{\m\n} + F_{\m\n} = \bar\th^{-1}_{\m\n} + \pa_\m A_\n - \pa_\n A_\m \,,\nn\\ 
 \d_\phi g_{\m\n}&=\d\phi^i \phi^j \eta_{ij} + \phi^i \d\phi^j \eta_{ij}\,, \nn\\
 \d_A F_{\m\n} &= \pa_\m \d A_\n - \pa_\n \d A_\m \,, \nn\\
h_{\m\n} &= -e^{\bar\s} (\bar\th^{-1} g F)_{\m\n} - e^{\bar\s} (F g \bar\th^{-1})_{\m\n} - \inv2 g_{\m\n}(\bar\th F)  + \cO(F^2) \,,
\end{align}
where $\bar\th^{-1}_{\m\n}$ denotes a self-dual Poisson structure with respect to a given metric $g_{\m\n}(\phi^i)$.
It was in fact argued in~\cite{Blaschke:2010qj}, that the tree level action \eqnref{YM--model} should single out almost self-dual geometries and that certain potential terms set the non-commutativity scale $e^\s\approx\,$const.

In the following section, we will consider two examples of geometries which are expected to solve the e.o.m. of the effective matrix model (i.e. including higher order contributions) too a good approximation~\cite{Blaschke:2010ye}, at least at some distance from the horizons.

\section{Special Geometries}
\subsection[Schwarzschild]{Schwarzschild Geometry}
%%%%%%%%%%%%%%%%%%%%%%%%%%%%%%%%%%%%%%%%%%%%%%
We now continue with the special example of Schwarzschild geometry, and our construction involves two steps~\cite{Blaschke:2010ye,Steinacker:2011yx}:

First, the choice of a suitable embedding $\cM^4\subset \R^D$ must be made 
such that the induced geometry on $\cM^4$ given by $g_{\mu\nu}$ is the Schwarzschild metric, and
then on needs to find a suitable non-degenerate Poisson structure on $\cM^4$ which solves the e.o.m. $\nabla^\mu \theta^{-1}_{\mu\nu} = 0$ for self-dual symplectic form $\syf$.
Both steps are far from unique a priori. However, the freedom is considerably reduced by 
requiring that the solution should be a ``local perturbation'' of an asymptotically flat
(or nearly flat) ``cosmological'' background. This is clear on physical grounds, having in mind
the geometry near a star in some larger cosmological context: It must be possible to 
approximately ``superimpose'' our solution, allowing e.g. for systems of stars and galaxies
in a natural way. This eliminates the well-known embeddings of the Schwarzschild geometry in the 
literature \cite{Kasner:1921,Fronsdal:1959zza,Kerner:2008nt}, which are highly non-trivial for large $r$ and cannot be superimposed in any obvious way.

Furthermore, the embedding should be asymptotically harmonic $\Box x^a \to 0$
for $r \to \infty$, in view of the fact that there may be terms in the matrix model which depend
on the extrinsic geometry, and which typically single out such harmonic 
embeddings\footnote{This can hold only asymptotically, since Ricci-flat geometries 
can in general \emph{not} be embedded harmonically \cite{Nielsen:1987}.}. Additionally, $\theta^{\mu\nu}$ should be non-degenerate, and $\th^{\mu\nu} \to \textrm{const.} \neq 0$ as $r \to \infty$.

We start by considering Eddington-Finkelstein coordinates  and define:
\begin{align}
t &= t_S + (r^* - r)\,,  
&
r^* &= r + r_c\ln\left|\frac{r}{r_c} - 1\right|
\,,
\end{align}
where $t_S$ denotes the usual Schwarzschild time, $r_c$ is the horizon of the Schwarzschild black hole and $r^*$ is 
the well-known tortoise coordinate. The metric in Eddington-Finkelstein coordinates $\{t,r,\vth,\vph\}$ is given by
\begin{align}
 ds^2 &= - \left( 1-\frac{r_c}{r} \right) dt^2 + \frac{2 r_c}{r}
    dt dr + \left( 1 + \frac{r_c}{r} \right) dr^2 + r^2 d\Omega^2 
\,, \label{eq:eddington-finkelstein-metric}
\end{align}
which is asymptotically flat for large $r$, and manifestly regular at the horizon $r_c$ and thus allows us to find an embedding which fulfills the requirements listed above. 
\begin{figure}[!ht]
\centering
\vspace{0.5cm}
\includegraphics[scale=0.9]{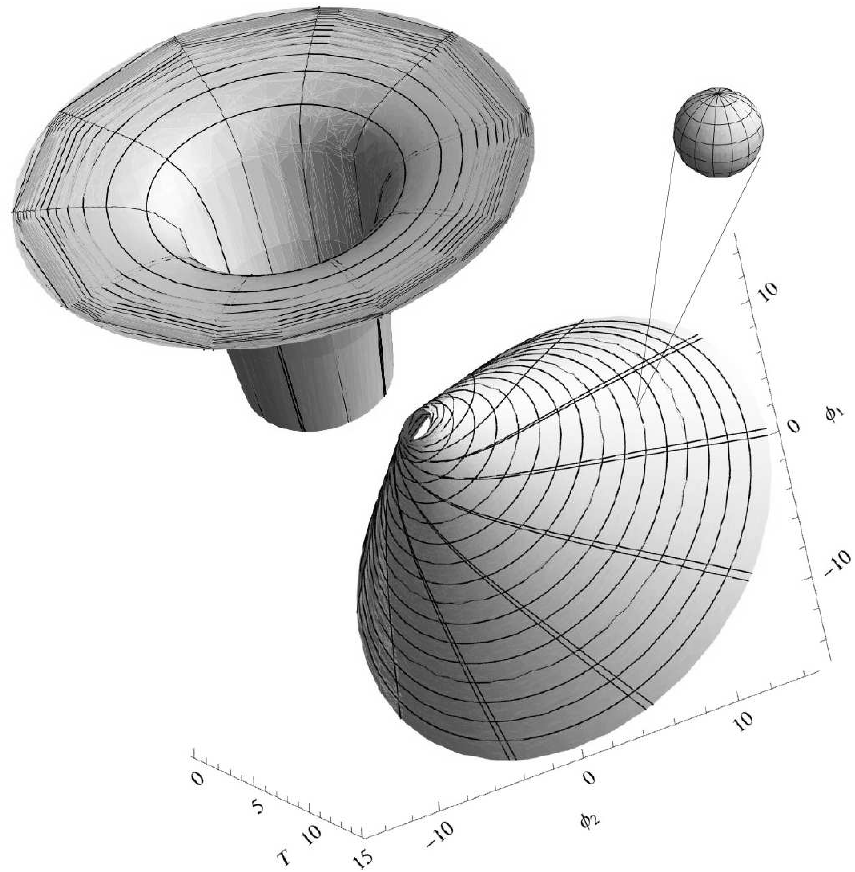}
\caption{Embedded Schwarzschild black hole.}
\label{fig:black-hole}
\end{figure}
In particular, we need at least 3 extra dimensions:
\begin{align}
 \phi_1+i\phi_2&=\phi_3e^{i\w (t + r)}\,, &
\phi_3&=\inv{\w}\sqrt{\frac{r_c}{r}}\,,
\end{align}
where $\phi_3$ is time-like and $\w$ is some parameter which does not enter the metric \eqref{eq:eddington-finkelstein-metric}. Hence, our 7-dimensional embedding is given by
\begin{align}
x^a&=\left(\begin{array}{c}
 t  \\
 r\cos\vph\sin\vth \\
 r\sin\vph\sin\vth \\
 r\cos\vth \\
 \inv{\w}\sqrt{\frac{r_c}{r}}\cos\left(\w(t+r)\right) \\[0.3em]
 \inv{\w}\sqrt{\frac{r_c}{r}}\sin\left(\w(t+r)\right) \\[0.3em]
 \inv{\w}\sqrt{\frac{r_c}{r}}
\end{array}\right)
\end{align}
with background metric $\eta_{ab}=\diag(-,+,+,+,+,+,-)$.

On the top of \figref{fig:black-hole}, a schematic view of the outer region of the 
Schwarzschild black hole is shown. After passing through the horizon $r=r_c$, the extra dimensions $\phi_i$ ``blow up'' 
in a cone-like manner. As indicated in the lower half of this figure, every point of the cone is in fact a sphere whose 
radius $r$ becomes smaller towards the bottom of the cone (i.e. $T\propto1/\sqrt{r}$). The twisted vertical lines drawn 
in the cone are lines of equal time $t$.

For the symplectic form, we require $\star\syf=i\syf$, so that the effective and the induced metric coincide, i.e. $G^{\m\n}=e^\s\th^{\m\r}\th^{\n\s}g_{\r\s}=g^{\m\n}$, and $\lim\limits_{r\to\infty}e^{-\s}=\textrm{const.}\neq0$. One then finds the solution~\cite{Blaschke:2010ye}
\begin{align}
\syf &= i E\wedge dt_S+B\wedge d\vph\,,\nn\\
E &= c_1 \left( \cos\vth dr -r\g\sin\vth d\vth\right)
 = d(f(r)\cos\vth)\,,\nn\\
B &= c_1 \left(r^2\sin\vth \cos\vth d\vth + r \sin^2\vth dr\right)
= \frac{c_1 }{2} d(r^2\sin^2\vth)\,,\nn\\
\g&=\left(1-\frac{r_c}{r}\right)\,,\qquad f(r)=c_1  r\g\,, \qquad f' = c_1 =\textrm{const.}\,,
\label{eq:schwarzschild-theta-sol}
\end{align}
from which follows
\begin{align}
e^{-\s}&=c_1^2\left(1-\frac{r_c}{r}\sin^2\vth\right)\equiv c_1^2e^{-\bar\s}
\,, 
\end{align}
where $c_1$ is an arbitrary constant. 
Clearly, near the horizon $e^{-\s} \approx \,$const. is not fulfilled, meaning our approximations break down in that region. Asymptotically, however, this seems to be a valid solution which approximately fulfills all requirements listed at the beginning of this section.

Furthermore, \eqnref{eq:schwarzschild-theta-sol} suggests to work in Darboux coordinates $x_D^\m=\{H_{ts},t_S,H_\vph,\vph\}$ corresponding to Killing vector fields $V_{ts}=\pa_{t_s}\,,\, V_\vph=\pa_\vph$ where the symplectic form $\syf$ is constant:
\begin{align}
\syf &= ic_1 d H_{ts} \wedge dt_S + c_1 d H_\vph \wedge d\vph\,, \nn\\
&=c_1 d\left(iH_{ts}dt_S+H_\vph d\vph\right)\,,\nn\\
H_{ts} &= r\g\cos\vth\,,
\qquad H_\vph = \inv{2} r^2\sin^2\vth
\,.
\end{align}
The relations to the Killing vector fields are given by
\begin{align}
E &= c_1  d H_{ts}  = c_1  E_\m dx^\m = i_{V_{ts}} \syf \,,   && E_\m = V_{ts}^\n \th^{-1}_{\n\m}\,,   \nn\\
B &= c_1  d H_\vph = c_1  B_\m dx^\m = i_{V_\vph} \syf \,,  && B_\m = V_\vph^\n \th^{-1}_{\n\m}\,,
\end{align}
and the metric in Darboux coordinates reads
\begin{align}
ds_{D}^2&=-\g dt_S^2+\frac{e^{\bar\s}}{\g}dH_{ts}^2+r^2\sin^2\vth d\vph^2+\frac{e^{\bar\s}}{r^2\sin^2\vth}dH_\vph^2
\,. \label{eq:ss-metric-darboux}
\end{align}
Hence, a Moyal type star product can easily be defined as
\begin{align}
(g\star h)(x_D) &= g(x_D) e^{-\frac{i}{2}\left(\lpa_\m\th_{D}^{\m\n}\rpa_\n\right)} h(x_D) \,,
&\textrm{with }\quad
\th_{D}^{\m\n}&=\e\left(\begin{array}{cccc}
0 & i & 0 & 0\\
-i & 0 & 0 & 0\\
0 & 0 & 0 & 1\\
0 & 0 & -1 & 0
\end{array}\right)\,,
\end{align}
where $\e=1/c_1 \ll1$ denotes the expansion parameter. 
Transforming back to embedding coordinates, the star product reads
\begin{align}
(g\star h)(x) &= g(x) \exp\Bigg[\frac{\ri\e}{2}\Bigg(\!\!\left(\lpa_t\frac{\ri r_cz e^{\bar\s}}{r^2\g}+\lpa_z \ri e^{\bar\s}\right)\wedge\rpa_t \nn\\*
&\quad +\!\left(\!\left(\lpa_t\!-\!\lpa_z\frac{z}{r}\right)\!\frac{r_c e^{\bar\s}}{r^2}\!+\!\left(\lpa_xx+\lpa_yy\right)\!\inv{x^2+y^2}\right)\!\wedge\!\left(x\rpa_y-y\rpa_x\right)\!\!\Bigg)\!\Bigg] h(x)
\end{align}
where the wedge stands for ``antisymmetrized'', and when considering the expansion 
care must be taken with the sequence of operators and the side they act on.
To leading order one hence finds the star commutators
\begin{align}
-\ri\starco{x^a}{x^b} &= \e e^{\bar\s} 
 \left(\!\!\!\begin{array}{ccccccc}
 0 & -\frac{r_cy}{r^2} & \frac{r_cx}{r^2} & -\ri & \frac{\ri zf^+_{12}(1)}{r} & \frac{\ri zf^-_{21}(1)}{r} & \frac{\ri z\phi_3}{2r^2} \\
 \frac{r_cy}{r^2} & 0 & e^{-\bar\s} & -\frac{r_cyz}{r^3}   & \frac{-yf^+_{12}(\g)}{r} & \frac{-yf^-_{21}(\g)}{r} & -\frac{y\g \phi_3}{2r^2}  \\
 -\frac{r_cx}{r^2} & -e^{-\bar\s} & 0 & \frac{r_cxz}{r^3} & \frac{xf^+_{12}(\g)}{r} & \frac{xf^-_{21}(\g)}{r} & \frac{x\g \phi_3}{2r^2} \\
 \ri & \frac{r_cyz}{r^3} & -\frac{r_cxz}{r^3} & 0 & -i\w\phi_2 & \ri\w\phi_1 & 0 \\
\!\!\frac{-\ri zf^+_{12}(1)}{r} & \frac{yf^+_{12}(\g)}{r} & \frac{-xf^+_{12}(\g)}{r} & i\w\phi_2 & 0 & -\frac{i\w z\phi_3^2}{2r^2} & \frac{-i\w z\phi_3\phi_2}{2r^2} \\
\!\!\frac{-\ri zf^-_{21}(1)}{r} & \frac{yf^-_{21}(\g)}{r} & \frac{-xf^-_{21}(\g)}{r} & -i \w\phi_1 & \frac{i\w z\phi_3^2}{2r^2} & 0 & \frac{i\w z\phi_3\phi_1}{2r^2} \\
 -\frac{\ri z\phi_3}{2r^2} & \frac{y\g \phi_3}{2r^2} & -\frac{x\g \phi_3}{2r^2} & 0 & \frac{\ri\w z\phi_3\phi_2}{2r^2} & \frac{-\ri\w z\phi_3\phi_1}{2r^2} & 0
\end{array}\!\!\!\!\!\right)
\nn\\
&\quad +\cO(\e^3)\,,
\end{align}
where
\begin{align}
f^{\pm}_{ij}(Y)&=\left(\frac{Y}{2r}\phi_i\pm\w\phi_j\right)
\,.
\end{align}
This defines a Poisson structure on $\cM^4$, but it could also be viewed as a Poisson structure
on the 6-dimensional space defined by $\phi_1^2 + \phi_2^2 = \phi_3^2$ which admits $\cM^4$ as symplectic leaf.

Higher orders in this star product, however, lead to {\nc} corrections to the embedding geometry, such as
$\phi_1\star\phi_1+\phi_2\star\phi_2 \neq \phi_3\star\phi_3$.

\subsection{Reissner-Nordstr{\"o}m Geometry}
%%%%%%%%%%%%%%%%%%%%%%%%%%%%%%%%%%%%%%%%%%%%%%
Similar to the Schwarzschild case, one can find an embedding with self-dual symplectic form also for the Reissner-Nordstr{\"o}m metric. In spherical coordinates $x^\m=\{t,r,\vth,\vph\}$ the according line element reads
\begin{align}
ds^2=-\left(1-\frac{2m}{r}+\frac{q^2}{r^2}\right)d\Tt^2+\left(1-\frac{2m}{r}+\frac{q^2}{r^2}\right)^{-1}dr^2+r^2d\W
\,. 
\end{align}
The geometry has two concentric horizons at
\begin{align}
r_h&=\left(m\pm\sqrt{m^2-q^2}\right)
\,. 
\end{align}
Shifting the time-coordinate according to
\begin{align}
t=\Tt+(r^*-r)\,, \qquad \textrm{with }dr^*\equiv \left(1-\tfrac{2m}{r}+\tfrac{q^2}{r^2}\right)^{-1}dr\,,
\end{align}
one arrives at
\begin{align}
ds^2&=-\left(1-\frac{2m}{r}+\frac{q^2}{r^2}\right)dt^2+2\left(\frac{2m}{r}-\frac{q^2}{r^2}\right)dtdr
 +\left(1+\frac{2m}{r}-\frac{q^2}{r^2}\right)dr^2+r^2d\W
\,. \label{eq:RN-metric-smooth}
\end{align}
We choose a 10-dimensional embedding $\cM^{1,3}\hookrightarrow\R^{4,6}$ which has the advantage of having similar properties compared to the Schwarzschild case\footnote{This choice, of course, is far from unique. Alternatively, we could have used a 7-dimensional embedding, but which would have been valid only up to the inner horizon. In fact, all physically relevant geometries should be embeddable in 10-dimensions, at least locally~\cite{Friedman:1961}.}. The additional coordinates $\phi_i$ are given by
\begin{align}
 \phi_1+i\phi_2&=\phi_3e^{i\w (t + r)}\,,\qquad
&\phi_3&=\inv{\w}\sqrt{\frac{2m}{r}}\,,\nonumber\\
\phi_4+i\phi_5&=\phi_6e^{i\w (t+r)}\,,\qquad
&\phi_6&=\frac{q}{\w r}
\,,
\end{align}
where $\phi_3$, $\phi_4$ and $\phi_5$ are \emph{time-like} coordinates. Like in the previous case, $\w$ does not enter the induced metric \eqref{eq:RN-metric-smooth}, but is hidden in the extra dimensions $\phi_i$. 
For $r\to\infty$, the $\phi_i$ become infinitesimally small and hence asymptotically, 
the four dimensional subspace becomes flat Minkowski space-time. 

An according self-dual symplectic form can be derived which in metric compatible Darboux coordinates reads
\begin{align}
\syf&=\inv{\e}\left(idH_{\Tt}\wedge d\Tt+dH_\vph\wedge d\vph\right)\,,\nonumber\\
H_{\Tt}&=\g\, r\cos\vth\,,\qquad\qquad
H_\vph=\frac{r^2}{2}\left(1-\frac{q^2}{r^2}\right)\sin^2\vth\,,\nonumber\\
\g&= \left(1-\frac{2m}{r}+\frac{q^2}{r^2}\right)
\,.
\end{align}
The non-commutativity scale in the outer region (i.e. at some distance to the horizon) is given by
\begin{align}
e^{-\bs}&=\g\sin^2\vth+\left(1-\frac{q^2}{r^2}\right)^2\cos^2\vth
\,, 
\end{align}
and the Reissner-Nordstr{\"o}m line element in Darboux coordinates reads
\begin{align}
ds_D^2=-\g d\Tt^2+\frac{e^\bs}{\g}dH_{\Tt}^2+r^2\sin^2\vth d\vph^2+\frac{e^\bs}{r^2\sin^2\vth}dH_\vph^2
\,, 
\end{align}
a form similar to the according Schwarzschild metric \eqref{eq:ss-metric-darboux}. 
In the limit $q\to0$ these expressions reduce to those in the Schwarzschild case\footnote{Note, that 3 of the extra dimensions, namely $\phi_{4-6}$ reduce to a point in this limit since $\lim\limits_{q\to0}\phi_6(q)=0$.}. 
Once more, a Moyal type star product can be defined as
\begin{align}
(g\star h)(x_D) &= g(x_D) e^{-\frac{\ri}{2}\left(\lpa_\m\th_{D}^{\m\n}\rpa_\n\right)} h(x_D) \,,
\end{align}
with the same block-diagonal $\th^{\m\n}$ as before. Higher orders in this star product lead to {\nc} corrections to the embedding geometry, such as $\phi_1\star\phi_1+\phi_2\star\phi_2 \neq \phi_3\star\phi_3$ and $\phi_4\star\phi_4+\phi_5\star\phi_5 \neq \phi_6\star\phi_6$ (see~\cite{Blaschke:2010ye} for details).

\section{Outlook}
%%%%%%%%%%%%%%%%%%%%%%%%%%%%%%%%%%%%%%
In this short proceeding note, explicit embeddings of Schwarzschild and Reissner-Nordstr{\"o}m geometries including self-dual symplectic forms have been discussed in the context of approximative solutions to the e.o.m. of an effective matrix model of Yang-Mills type. 
It was pointed out, that in a future effort the embeddings should be modified near the horizons to account for nearly constant $e^\s$. 
Open questions, among others, concern deviations from $G=g$ and higher order quantum effects.

\subsection*{Acknowledgements}
%%%%%%%%%%%%%%%%%%%%%%%%%%%%%%%%%%%%%%%%%%
Many thanks go to the organizers of the 2010 workshop on non-commutative field theory and gravity in Corfu, which was a wonderful and stimulating conference. 
This work was supported by the Austrian Science Fund (FWF) under contract P21610-N16.